\documentclass{article}
\usepackage{times}
\usepackage{amsfonts}
\usepackage{graphicx}
\usepackage[pdfmark]{hyperref}
\begin{document}
\noindent
{\Large  DIRAC BRACKETS FROM MAGNETIC BACKGROUNDS}
\vskip1cm
\noindent
{\bf Jos\'e M. Isidro}\\
Instituto de F\'{\i}sica Corpuscular (CSIC--UVEG), Apartado de Correos 22085,\\ Valencia 46071, Spain\\
{\tt jmisidro@ific.uv.es}
\vskip1cm

\noindent
{\bf Abstract} 
\noindent
In symplectic mechanics, the magnetic term describing the interaction between a charged particle and an external magnetic field has to be introduced by hand. On the contrary, in  generalised complex geometry, such magnetic terms in the symplectic form arise naturally by means of $B$--transformations. Here we prove that, regarding classical phase space as a generalised complex manifold, the transformation law for the symplectic form under the action of a weak magnetic field gives rise to Dirac's prescription for Poisson brackets in the presence of constraints. 


\section{Introduction}\label{rmllgurr}

Let $\mathbb{P}$ denote a $2n$--dimensional symplectic manifold, that we take to be the classical phase space of a system with $n$ independent degrees of freedom.  
In local coordinates $x^1,\ldots,x^{2n}$ on $\mathbb{P}$, let the symplectic form $\omega$ and the Poisson brackets $\{\cdot,\cdot\}$ be given by
\begin{equation}
\omega=\frac{1}{2}\omega_{ij}{\rm d}x^i\wedge{\rm d}x^j, \qquad \left\{f,g\right\}=\pi^{ij}\partial_if\partial_jg,
\label{cvt}
\end{equation}
where $\omega_{ij}\pi^{jl}=\delta_i^l$, and $f,g\in C^{\infty}(\mathbb{P})$ are any two smooth functions. 

Dirac brackets were introduced as a modification of Poisson brackets in the presence of constraints \cite{DIRAC}; their definition can be summarised as follows. Assume imposing $2n'$ independent constraints, which are satisfied on a $2(n-n')$--dimensional symplectic submanifold $\mathbb{P'}\subset\mathbb{P}$. In the neighbourhood of a point $x'\in\mathbb{P'}$, choose coordinates $y^1,\ldots, y^{2n}$ on $\mathbb{P}$ such that $\mathbb{P'}$ is given by
\begin{equation}
y^{1}=0, \ldots, y^{2n'}=0,
\label{ctr}
\end{equation}
so that $y^{2n'+1}, \ldots, y^{2n}$ provide local coordinates on $\mathbb{P'}$. Next consider the matrix $C^{rs}(y)$ whose entries are defined by
\begin{equation}
C^{rs}(y):=\left\{y^{r},y^{s}\right\}, \qquad r,s=1,\ldots, 2n'.
\label{mmtr}
\end{equation}
{}Finally assume that the matrix $C^{rs}(y)$ is invertible at $x'$, and let $C_{rs}(y)$ denote its inverse. Given $f,g\in C^{\infty}(\mathbb{P})$, let $f', g'$ be their respective restrictions to $\mathbb{P'}$. Then Dirac brackets are defined by
\begin{equation}
\left\{f', g'\right\}_{\rm Dirac}:=\left\{f,g\right\}-\sum_{r,s=1}^{2n'}\left\{f,y^r\right\}C_{rs}\left\{y^{s},g\right\}.
\label{dbr}
\end{equation}

Along an apparently unrelated line, physics in the presence of magnetic backgrounds has been the subject of intense recent research using the tools of noncommutative geometry \cite{CONNES, LANDI, SZABO}. Here the archetypal theory is the Landau model: electric charges on the $x,y$ plane subject to an external magnetic field ${\bf B}$ applied perpendicularly to the plane. Then the canonical momenta no longer Poisson--commute \cite{LANDAU}: 
\begin{equation}
\left\{p_x,p_y\right\}=\frac{e\vert {\bf B}\vert}{c}.
\label{rmlcps}
\end{equation}
Noncommutative geometry, by now popular among physicists especially after the works of refs. \cite{CDS, SW}, is however not the only arena to study physics in the presence of magnetic backgrounds. Regarding the right--hand side of eqn. (\ref{rmlcps}) as a central extension, it can be studied using cohomological methods \cite{AZCARRAGA}. Last but not least, recent breakthroughs in the theory of symplectic and complex manifolds, that go by the name of {\it generalised complex structures}\/ \cite{HITCHIN, GUALTIERI}, provide an interesting alternative to analyse physics in the presence of magnetic fields. Thus, {\it e.g.}, in ref. \cite{MAGNETIC} we have used the techniques of generalised complex geometry in order to prove that the Poisson brackets of the Landau model have a natural origin within the framework of generalised complex manifolds; further applications have been studied in ref. \cite{IJMPA}. Given the major role that symplectic geometry plays in the mechanics of classical and quantum systems \cite{GOSSON}, one can expect generalised complex geometry to become an essential tool, especially within the context of phase--space quantum mechanics \cite{ZACHOS, NOI}. In this latter framework, Dirac brackets have been the subject of recent attention \cite{KRIVORUCHENKO}.

One can in fact regard an external magnetic field as a constraint: ${\bf B}\neq 0$ induces the nonvanishing of the right--hand side of (\ref{rmlcps}). The resulting equation qualifies as a constraint in Dirac's sense, while the usual $\left\{p_x,p_y\right\}=0$ does not. In this paper we prove that the converse is also true: given a set of Dirac constraints (\ref{ctr}), we derive a magnetic field that induces the corresponding Dirac brackets (\ref{dbr}). We will see that the latter have a natural origin in terms of generalised complex geometry. By extending our understanding of classical phase space and regarding it as a generalised complex manifold, a unified picture is obtained in which magnetic fields and Dirac constraints turn out to be equivalent objects. Our approach may be understood as complementary to that provided by noncommutative geometry. Indeed, in the latter case one often takes the limit of strong magnetic fields, whereas our analysis here will concentrate on the limit of weak magnetic fields.

Our conventions are as follows. Omitting the indices from $\pi^{ij}$, we will denote the Poisson tensor by $\pi$. We will employ Einstein's convention when summing over all $2n$ dimensions of $\mathbb{P}$, as in eqn. (\ref{cvt}). However, sums over the $2n'$ dimensions of $\mathbb{P}-\mathbb{P'}$ will be explicitly indicated, as in eqn. (\ref{dbr}).

\section{Phase space as a generalised complex manifold}\label{bbnktdpklmrkdmrd}

\subsection{Basics in generalised complex structures}\label{csgmzktjd}

Let $\mathbb{M}$ be a smooth manifold of dimension $2n$. We will illustrate our statements in local coordinates around a point $x\in\mathbb{M}$, forgetting about global issues that can be taken care of by the appropriate integrability conditions. For details we refer the reader to \cite{GUALTIERI}.  

The total space of the bundle $T\mathbb{M}\oplus T^*\mathbb{M}$ is real $6n$--dimensional: $2n$ dimensions for the base, $4n$ for the fibre. A {\it generalised complex structure}\/ over $\mathbb{M}$, denoted ${\cal J}$, is an endomorphism of the fibre over each $x\in\mathbb{M}$,
\begin{equation}
{\cal J}_x\colon T_x\mathbb{M}\oplus T^*_x\mathbb{M}\longrightarrow T_x\mathbb{M}\oplus T_x^*\mathbb{M},
\label{rmlltnsugjrpkl}
\end{equation}
satisfying the following three conditions.  First, for all $x\in \mathbb{M}$ one has 
\begin{equation}
{\cal J}_x^2=-{\bf 1}.
\label{rmllkthst}
\end{equation}
Second, for all $x\in\mathbb{M}$ one has 
\begin{equation}
{\cal J}_x^t=-{\cal J}_x,
\label{brbkttlspelskttcsp}
\end{equation}
the superindex ${}^t$ standing for transposition. Third, the {\it Courant integrability condition}\/ must hold; in what follows we will assume that this latter condition is always satisfied. Generalised complex geometry thus involves an object ${\cal J}$ that is simultaneously complex, by eqn. (\ref{rmllkthst}), and symplectic, by eqn. (\ref{brbkttlspelskttcsp}). 

Suppose that ${\cal J}$ at $x\in\mathbb{M}$ is given by
\begin{equation}
{\cal J}_{\omega_x}:=\left(\begin{array}{cc}
0&-\omega_x^{-1}\\
\omega_x&0\end{array}\right),
\label{ramllkomprtchantikspa}
\end{equation}
$\omega$ being a symplectic form and $\omega_x$ its evaluation at $x\in\mathbb{M}$; the decomposition into block matrices reflects the direct sum (\ref{rmlltnsugjrpkl}). This ${\cal J}_{\omega}$ defines a generalised complex structure {\it of symplectic type}\/ on $\mathbb{M}$. At the other end we have that
\begin{equation}
{\cal J}_{J_x}:=\left(\begin{array}{cc}
-J_x&0\\
0&J^t_x\end{array}\right),
\label{rmllotenllenksaps}
\end{equation}
$J$ being a complex structure and $J_x$ its evaluation at $x\in\mathbb{M}$, defines a generalised complex structure {\it of complex type}\/ on $\mathbb{M}$. 

There exists a Darboux--like theorem describing the local form of a generalised complex structure in the neighbourhood of any regular point $x\in\mathbb{M}$. Roughly speaking, any manifold endowed with a generalised complex structure splits {\it locally}\/ as the product of a complex manifold times a symplectic manifold. A more precise statement is as follows. A point $x\in\mathbb{M}$ is said {\it regular}\/ if the Poisson structure $\omega^{-1}$ has constant rank in a neighbourhood of $x$. Then any regular point in a generalised complex manifold has a neighbourhood which is equivalent, via a diffeomorphism and a $B$--transformation, to the product of an open set in $\mathbb{C}^k$ and an open set in $\mathbb{R}^{2n-2k}$, the latter endowed with its standard symplectic form. The nonnegative integer $k$ is called the {\it type}\/ of ${\cal J}$, $k=0$ and $k=n$ being the limiting cases examined in eqns. (\ref{ramllkomprtchantikspa}) and (\ref{rmllotenllenksaps}), respectively.

Next assume that $\mathbb{M}$ is a linear space. Then any generalised complex structure of type $k=0$ is the $B$--transform of a symplectic structure. This means that any generalised complex structure of type $k=0$ can be written as
$$
{\rm e}^{-B}{\cal J}_{\omega}{\rm e}^{B}=
\left(\begin{array}{cc}
{\bf 1}&0\\
-B&{\bf 1}\end{array}\right)
\left(\begin{array}{cc}
0&-\omega^{-1}\\
\omega &0\end{array}\right)
\left(\begin{array}{cc}
{\bf 1}&0\\
B&1\end{array}\right)
$$
\begin{equation}
=\left(\begin{array}{cc}
-\omega^{-1}B & -\omega^{-1}\\
\omega + B\omega^{-1}B & B\omega^{-1}\end{array}\right),
\label{rmlstsllncspass}
\end{equation}
for a certain 2--form $B$. Similarly any generalised complex structure of type $k=n$ over a linear manifold $\mathbb{M}$ is the $B$--transform of a complex structure,
$$
{\rm e}^{-B}{\cal J}_{J}{\rm e}^{B}=
\left(\begin{array}{cc}
{\bf 1}&0\\
-B&{\bf 1}\end{array}\right)
\left(\begin{array}{cc}
-J&0\\
0 & J^t\end{array}\right)
\left(\begin{array}{cc}
{\bf 1}&0\\
B&{\bf 1}\end{array}\right)
$$
\begin{equation}
=\left(\begin{array}{cc}
-J & 0\\
BJ+J^tB & J^t\end{array}\right).
\label{rmlldchtkehuelesmrd}
\end{equation}
When $\mathbb{M}$ is an arbitrary smooth manifold, not necessarily a linear space, the statements around eqns. (\ref{rmlstsllncspass}) and (\ref{rmlldchtkehuelesmrd}) hold essentially true, with some minor refinements required; see ref. \cite{GUALTIERI} for details.

\subsection{The metric}\label{luisist}

In what follows we take $\mathbb{M}$ to be the classical phase space $\mathbb{P}$ of section \ref{rmllgurr}. Since the latter is symplectic it automatically qualifies as generalised complex of type $k=0$. Let us see that a metric of signature $(2n,2n)$ such as that of ref. \cite{GUALTIERI} can be readily manufactured on $\mathbb{P}$ with the mechanical elements at hand.

Let us pick Darboux coordinates $q^j, p_j$ around $x\in\mathbb{P}$, where $j=1,\ldots,n$.  Let an ordered basis for $T_x^*\mathbb{P}$ be spanned by 
\begin{equation}
{\rm d}q^1,\ldots, {\rm d}q^n, {\rm d}p_1, \ldots, {\rm d}p_n.
\label{lvzzrhj}
\end{equation}
Correspondingly, an ordered basis for $T_x\mathbb{P}\oplus T_x^*\mathbb{P}$ is spanned by 
\begin{equation}
\partial_{q^1},\ldots, \partial_{q^n},\partial_{p_1}, \ldots, \partial_{p_n}, {\rm d}q^1,\ldots, {\rm d}q^n,{\rm d}p_1, \ldots, {\rm d}p_n.
\label{laghijpt}
\end{equation}
Starting from the classical symplectic form $\omega$ in Darboux coordinates, its (block) matrix at $x\in\mathbb{P}$ is
\begin{equation}
\omega=\left(\begin{array}{cc}
0&{\bf 1}_n\\
-{\bf 1}_n&0\end{array}\right).
\label{bbnkthstpldrn}
\end{equation}
The Poisson tensor $\pi$ is the inverse of the symplectic matrix $\omega$,
\begin{equation}
\pi=\omega^{-1}.
\label{rmllmird}
\end{equation}
In Darboux coordinates we have $\omega^{-1}=-\omega$ and $(-\omega)^2=-{\bf 1}$. Now ${\rm i}\hbar$ times classical Poisson brackets are quantum commutators. Hence the latter are represented by 
\begin{equation}
{\rm i}\hbar\left(\begin{array}{cc}
0&-{\bf 1}_n\\
{\bf 1}_n&0\end{array}\right).
\label{bbrnkthstprmrkn}
\end{equation}
Setting $\hbar=1$, the above squares to the identity. The direct sum of the squares of the matrices (\ref{bbrnkthstprmrkn}), (\ref{bbnkthstpldrn}) gives us the expression, in Darboux coordinates, of a diagonal metric $G$ on $T\mathbb{P}\oplus T^*\mathbb{P}$
\begin{equation}
G=\left(\begin{array}{cc}
{\bf 1}_{2n}&0\\
0&-{\bf 1}_{2n}\end{array}\right).
\label{brbntsht}
\end{equation}
As explained in ref. \cite{IJMPA}, the fact that a metric of the required signature can be constructed from the (classical and quantum) mechanical elements present in $T\mathbb{P}\oplus T^*\mathbb{P}$ indicates that our mechanical setup can make contact, in a natural way, with the geometry of generalised complex manifolds, where the split--signature metric (\ref{brbntsht}) plays a major role \cite{GUALTIERI}.

\subsection{$B$--transformation of the Poisson brackets}\label{rmyktprtnry}

Under a $B$--transformation, the lower left entry of ${\cal J}_{\omega}$ in  eqn. (\ref{rmlstsllncspass}) transforms as
\begin{equation}
\omega\longrightarrow\omega_B:=\omega+B\omega^{-1}B=\omega+B\pi B.
\label{rmllokk}
\end{equation}
We will see presently that $\omega_B$ qualifies as a symplectic form. Consider the generalised complex structure ${\cal J}_{\omega_B}$ of type $k=0$ defined on $\mathbb{P}$ by
\begin{equation}
{\cal J}_{\omega_B}:=\left(\begin{array}{cc}
0&-\omega_B^{-1}\\
\omega_B&0\end{array}\right).
\label{lagmkgntptmdr}
\end{equation}
Then we have a $B$--transform $\pi_B$ of the Poisson tensor $\pi$,
\begin{equation}
\pi_B:=\omega_B^{-1}=(\omega+B\omega^{-1}B)^{-1}.
\label{ptrmall}
\end{equation}
If $B$ is sufficiently weak, we can neglect terms of order $O(B^4)$ and higher. In this way $\pi_B$ can be approximated by
\begin{equation}
\pi_B=\pi-\pi B\pi B\pi+O(B^4),
\label{trn}
\end{equation}
because then
\begin{equation}
\pi_B\omega_B=1+O(B^4), 
\label{yldr}
\end{equation}
as is readily verified. In what remains we will always work in the weak--field approximation.

\section{Dirac constraints as weak magnetic fields}\label{brbnmkgntsptsmrts}

By eqn. (\ref{trn}), the weak--field Poisson brackets of $f,g$ are
\begin{equation}
\{f,g\}_B=\left(\pi_B\right)^{ij}\partial_if\partial_jg=\{f,g\}-\left(\pi B\pi B\pi\right)^{ij}\partial_if\partial_jg.
\label{ptnrmllno}
\end{equation}
Comparing the above with eqn. (\ref{dbr}), the following identification is suggested:
\begin{equation}
\pi_{\rm Dirac}=\pi_B.
\label{csrgomzrsnmprstbldmrda}
\end{equation}
This is equivalent to trading the initial set of constraints (\ref{ctr}) for a weak magnetic field $B$ such that the following condition holds:
\begin{equation}
\left (\pi B\pi B\pi\right)^{ij}\partial_if\partial_jg=\sum_{r,s=1}^{2n'}\left\{f,y^r\right\}C_{rs}\left\{y^s,g\right\}.
\label{rmllhyklvrslsmnsdspdkgr}
\end{equation}
We can use the above to solve for $B$. By eqn. (\ref{cvt}) 
\begin{equation}
\left(\pi B\pi B\pi\right)^{ij}=\sum_{r,s=1}^{2n'}\pi^{il}\partial_{l}y^r\,C_{rs}\,\partial_{m}y^s\pi^{mj},
\label{mrtrzdcrls}
\end{equation}
which, given the nondegeneracy of $\pi$, is equivalent to
\begin{equation}
\left(B\pi B\right)_{ij}=\sum_{r,s=1}^{2n'}\partial_{i}y^r\,C_{rs}\,\partial_{j}y^s.
\label{btrzmrtrzzz}
\end{equation}
Eqn. (\ref{btrzmrtrzzz}) determines $B$ as a function of the constraints. The weak--field approximation made here means that quadratic terms in the magnetic field $B$ correspond to linear terms in the (inverse) matrix of constraints $C$:
\begin{equation}
O(B^2)\approx O(C).
\label{brtcklosmrtz}
\end{equation}

The above equivalence between weak magnetic fields and Dirac constraints can also be derived alternatively as follows.  By eqn. (\ref{dbr}) we  have
\begin{equation}
\{f,g\}_{\rm Dirac}=\pi_{\rm Dirac}^{ij}\partial_if\partial_jg=\left(\pi^{ij}-\sum_{r,s=1}^{2n'}\pi^{il}\partial_{l}y^r\,C_{rs}\,\partial_{m}y^s\pi^{mj}\right)\partial_if\partial_jg.
\label{brtkgptmrt}
\end{equation}
Given that the constrained manifold $\mathbb{P'}$ is assumed symplectic, we can also write
\begin{equation}
\{f,g\}_{\rm Dirac}=\left(\omega_{\rm Dirac}^{-1}\right)^{ij}\partial_if\partial_jg.
\label{btrzptz}
\end{equation}
Thus equating (\ref{brtkgptmrt}) and (\ref{btrzptz}) we obtain
\begin{equation}
\pi_{\rm Dirac}^{ij}=\left(\omega_{\rm Dirac}^{-1}\right)^{ij}=\pi^{ij}-\sum_{r,s=1}^{2n'}\pi^{il}\partial_{l}y^r\,C_{rs}\,\partial_{m}y^s\pi^{mj}.
\label{ptbrtdcrls}
\end{equation}
We can solve for $\omega_{\rm Dirac}$ to the same degree of approximation given by (\ref{brtcklosmrtz}):
\begin{equation}
\left(\omega_{\rm Dirac}\right)_{ij}=\omega_{ij}+\sum_{r,s=1}^{2n'}\partial_i y^r\,C_{rs}\,\partial_j y^s+O(C^2).
\label{brtzmrdxxx}
\end{equation}
Comparing eqns. (\ref{rmllokk}) and (\ref{brtzmrdxxx}) we recover  our previous result (\ref{btrzmrtrzzz}). The previous arguments also establish that $\omega_B$ qualifies as a symplectic form, because $\omega_B=\omega_{\rm Dirac}$.

Eqn. (\ref{btrzmrtrzzz}) can be made more precise as follows. In the Darboux coordinates of (\ref{bbnkthstpldrn}), assume a block decomposition for $B$
\begin{equation}
B=\left(\begin{array}{cc}
B_{qq}&B_{qp}\\
-B_{qp}^t&B_{pp}
\end{array}\right), \qquad B^t_{qq}=-B_{qq}, \quad \qquad B^t_{pp}=-B_{pp},
\label{gumkgntptmdr}
\end{equation}
where the $n\times n$ matrices $B_{qq}$, $B_{qp}$ and $B_{pp}$ are so chosen that $B$ is antisymmetric. Above, the $n$--valued indices $q,p$ run over the range specified in (\ref{lvzzrhj}). By eqns. (\ref{bbnkthstpldrn}) and (\ref{rmllmird}),
\begin{equation}
B\pi B=\left(\begin{array}{cc}
B_{qp}B_{qq}+B_{qq}B_{qp}^t&B_{qp}^2-B_{qq}B_{pp}\\
B_{pp}B_{qq}-(B_{qp}^t)^2&B_{pp}B_{qp}+B_{qp}^tB_{pp}
\end{array}\right).
\label{alvrzmsrblxxx}
\end{equation}
This must be equated to the block decomposition of $\sum_{r,s}\partial_iy^rC_{rs}\partial_jy^s$ in Darboux coordinates, where $i,j$ are Darboux indices. Assume that we have the block decomposition into $n\times n$ matrices
\begin{equation}
\sum_{r,s=1}^{2n'}\partial_iy^rC_{rs}\partial_jy^s=
\left(\begin{array}{cc}
Y_{qq}&Y_{qp}\\
-Y_{qp}^t&Y_{pp}\end{array}\right), 
\label{alvrzmsdmrddd}
\end{equation}
where the $n$--valued indices $q,p$ are as in (\ref{gumkgntptmdr}), and
\begin{equation}
Y^t_{qq}=-Y_{qq}, \quad  Y^t_{pp}=-Y_{pp}.
\label{juntoh}
\end{equation}
Equating the corresponding matrices of  (\ref{alvrzmsrblxxx}) and (\ref{alvrzmsdmrddd}) we obtain a system of $n\times n$ matrix relations determining the block entries of $B$ in Darboux coordinates. The resulting equations are not very illuminating. However they can be substantially simplified by the following observation: a symplectic transformation will reduce the right--hand side of (\ref{alvrzmsdmrddd}) to a form such that
\begin{equation}
Y_{qq}=0, \qquad Y_{pp}=0.
\label{vrzmsrbkj}
\end{equation}
{}For simplicity we will denote the new corodinates also by $q^j$, $p_j$. This is a symplectic transformation with respect to the symplectic structure on $\mathbb{P}$; it need not, and generally it will not, be a symplectic transformation with respect to the symplectic structure on $\mathbb{P}'$, as Darboux coordinates for the latter need not be Darboux coordinates for the former. The symplectic transformation around eqn. (\ref{vrzmsrbkj}) merely block--antidiagonalises eqn. (\ref{alvrzmsdmrddd}), but it need not, and generally it will not, reduce $Y_{qp}$ to the identity ${\bf 1}_n$. By (\ref{vrzmsrbkj}) we can now pick
\begin{equation}
B_{qq}=0, \qquad B_{pp}=0.
\label{kip}
\end{equation}
Now (\ref{vrzmsrbkj}) and (\ref{kip}) considerably simplify the equality between the right--hand sides of (\ref{alvrzmsrblxxx}) and (\ref{alvrzmsdmrddd}), which reduces to 
\begin{equation}
B_{qp}^2=Y_{qp}=\sum_{r,s=1}^{2n'}\partial_qy^r\,C_{rs}\,\partial_py^s.
\label{law}
\end{equation}
To summarise, in Darboux coordinates, the weak magnetic field that is equivalent to the initial Dirac constraints is given by
\begin{equation}
B=\pm b\left(\begin{array}{cc}
0&+\sqrt{\sum_{r,s=1}^{2n'}\partial_qy^r\,C_{rs}\,\partial_py^s}\\
-\sqrt{\sum_{r,s=1}^{2n'}\partial_py^r\,C_{rs}\,\partial_qy^s}&0
\end{array}\right),
\label{mkgkkdrr}
\end{equation}
where the lower--left entry is the transpose of the upper--right matrix. Above we have included an infinitesimal parameter $b\in\mathbb{R}$ to ensure the validity of the weak--field approximation; we can identify $b=\vert {\bf B}\vert$, the latter as in (\ref{rmlcps}).

\section{Discussion}\label{jvrmsimbcc}

Our main conclusion is that Dirac constraints are equivalent to weak magnetic fields, in the following sense. Regard classical phase space $\mathbb{P}$ as a generalised complex manifold of symplectic type, and apply a (weak) $B$--transformation. This is interpreted physically as the action of a weak magnetic field. Then
the transformation law for Poisson brackets on the unconstrained manifold $\mathbb{P}$  gives rise to Dirac's prescription for Poisson brackets on the constrained manifold $\mathbb{P}'$. In this way we can trade the Dirac brackets of {\it constrained}\/ functions $f', g'$ ({\it i.e.}, functions $f,g$ in which some slots have been filled in with $y^1=0,\ldots,y^{2n'}=0$) for the weak--field brackets of the same {\it unconstrained}\/ functions $f,g$. 
 
Weakness of the magnetic field is imposed by the requirement that the approximation $(1-z)^{-1}\approx 1+z$ hold.  Somehow, this limit also appears to have a physical counterpart in Dirac's own prescription (\ref{dbr}): the latter contains just one power of the (inverse) matrix of constraints $C$. However this apparent limitation to {\it weak}\/ magnetic fields is in fact a desirable feature of our approach, because of the following reason. The Fourier transform of eqn. (\ref{rmlcps}),
\begin{equation}
\left\{x,y\right\}=\frac{c}{e\vert {\bf B}\vert},
\label{btrzrr}
\end{equation}
becomes, upon quantisation, 
\begin{equation}
[X,Y]={\rm i}\hbar\theta, \qquad \theta:=\frac{c}{e\vert {\bf B}\vert},
\label{bbnmkgtdppmts}
\end{equation}
where we have introduced the usual noncommutativity parameter $\theta$ between operator coordinates $X$ and $Y$. Quantum theories on noncommutative spacetime are often analysed, using the tools of noncommutative geometry, in the limit $\theta\to 0$ \cite{LIZZI}. This latter limit corresponds to strong magnetic fields, while the case of a strong noncommutativity parameter $\theta\to\infty$ falls outside the range of applicability of such analyses. On the contrary, our approach via generalised complex structures is well suited to the case of weak magnetic fields.

An interesting geometric property to realise is that the $B$--transformation laws (\ref{rmllokk}) for the symplectic form  and (\ref{ptrmall}) for the Poisson tensor are nontensorial. In fact $B$--transformations are {\it not}\/ diffeomorphisms of $\mathbb{P}$, so it would be a surprise if $\omega$ and $\pi$ transformed tensorially under $B$. By regarding classical phase space as generalised complex, rather than just symplectic, the type $k$ of $\mathbb{P}$ remains zero. However one gains the advantage that one can apply the $B$--transformations which are interpreted physically as magnetic fields. This raises the following intriguing point. We know that gravitational fields are equivalent to accelerated frames; it was just proved here that weak magnetic fields are equivalent to Dirac constraints. What implications does this have on the equivalence principle of quantum mechanics \cite{MATONE}?
Our conclusions also contribute towards a modern geometric view of quantum mechanics,  a beautiful presentation of which has been given in ref. \cite{GMS}.

\noindent
{\bf Acknowledgements} It is a great pleasure to thank Albert--Einstein--Institut (Potsdam, Germany) for hospitality during the preparation of this article. This work has been supported by Ministerio de Educaci\'{o}n y Ciencia (Spain) through grant FIS2005--02761, by Generalitat Valenciana, by EU FEDER funds, by EU network MRTN--CT--2004--005104 ({\it Constituents, Fundamental Forces and Symmetries of the Universe}), and by Deutsche Forschungsgemeinschaft.

\end{document}